\documentclass[a4paper,11pt]{article}
\pdfoutput=1
\usepackage[dvipdfmx]{graphicx}
\usepackage{bm,bigdelim,multirow}
\usepackage{booktabs}
\usepackage{braket,latexsym}
\usepackage{jheppub}
\usepackage[T1]{fontenc} 
\usepackage[numbers]{natbib}
\newcommand{\Tr}{\mathrm{Tr}}
\newcommand{\e}{\mathrm{e}}
\newcommand{\T}{\mathrm{T}}
\newcommand{\Dcut}{D_{\mathrm{cut}}}
\newcommand{\Tc}{T_{\mathrm{c}}}

\begin{document}

  \title{Phase transition of four-dimensional Ising model with higher-order tensor renormalization group}

  \author[a]{Shinichiro Akiyama,}
  \emailAdd{akiyama@het.ph.tsukuba.ac.jp}

  \author[b]{Yoshinobu Kuramashi,}
  \emailAdd{kuramasi@het.ph.tsukuba.ac.jp}

 \author[c]{Takumi Yamashita,}
 \emailAdd{yamasita@ccs.tsukuba.ac.jp}
 
 \author[b]{and Yusuke Yoshimura}
 \emailAdd{yoshimur@ccs.tsukuba.ac.jp}

  \affiliation[a]{Graduate School of Pure and Applied Sciences, University of Tsukuba, Tsukuba, Ibaraki
    305-8571, Japan}
  \affiliation[b]{Center for Computational Sciences, University of Tsukuba, Tsukuba, Ibaraki
    305-8577, Japan}
  \affiliation[c]{Faculty of Engineering, Information and Systems, University of Tsukuba, Tsukuba, Ibaraki
    305-8573, Japan}  
    
 \abstract{We apply the higher-order tensor renormalization group (HOTRG) to the four-dimensional ferromagnetic Ising model, which has been attracting interests in the context of the triviality of the scalar $\phi^4_{d=4}$ theory. We investigate the phase transition of this model with HOTRG enlarging the lattice size up to $1024^4$ with parallel computation. The results for the internal energy and the magnetization are consistent with the weak first-order phase transition.}

\preprint{UTHEP-723, UTCCS-P-115}

\maketitle
\flushbottom

\section{Introduction}
\label{sec:intro}

It is well known that the critical behavior of the Ising model on the higher-dimensional hypercubic lattice is well explained with the mean-field theory. In dimensions larger than four, the effect of the background fluctuations becomes negligible and the model in the critical region exactly obeys the mean-field exponents \cite{Aizenman:1981zz,Aizenman:1982ze}.  At the upper critical dimension, however, multiplicative logarithmic corrections are added to the leading scaling behavior of the mean-field theory. Some of these corrections were derived by the perturbative calculation with the renormalization group method \cite{PhysRevB.7.248}. Since the Ising model is specified by the infinite coupling limit of the single-component scalar $\phi^4_4$ theory, the model in four dimensions has been attracting the interest of particle physicists for a long time in the context of the triviality of the $O(4)$ scalar $\phi^4_4$ theory, which corresponds to the scalar sector of the standard model describing the generation of gauge boson and fermion mass through the Higgs mechanism \cite{Wilson:1973jj,Luscher:1987ay,Luscher:1987ek,Luscher:1988uq,Huang:1988hu,Jansen:1988cw,Frick:1989gw}. There is also a recent study to discuss the triviality of the $O(N)~\phi^4_4$ theory with the higher-loop beta function \cite{Shrock:2014zca,Shrock:2016hqn,Shrock:2017zuk}.

A numerical study of the Ising model on a hypercubic lattice serves as a non-perturbative test of the triviality \cite{Kenna:1992np,Kenna:2004cm}; if the leading scaling behavior is the mean-field type and it is modified only by the multiplicative logarithmic factor, one obtains a supporting evidence for the triviality. In fact, the numerical investigation based on the Monte Carlo simulation has successfully caught the mean-field exponents \cite{PhysRevB.22.4481,SanchezVelasco:1987ah,Bittner:2002pk,PhysRevE.80.031104,Lundow:2010en}, but there remains some controversy over the appearance of the logarithmic corrections \cite{PhysRevE.80.031104,Lundow:2010en,Cea:2005ad,Stevenson:2005yn,Balog:2006fs}. Actually no Monte Carlo study has confirmed the logarithmic correction in the scaling behavior of the specific heat, which is $(\ln |t|)^{1/3}$ with $t$ the reduced temperature expected from the perturbative renormalization group analysis. This is mainly because the cubic root of logarithmic divergence is too weak to detect by the finite size scaling analysis, or the specific heat may be actually bounded \cite{PhysRevE.80.031104}. Indeed, the finite volume effect of the four-dimensional Ising model had been investigated from various viewpoints \cite{Jansen:1988cw}. A detailed Monte Carlo study has found serious finite-volume effect due to non-trivial boundary effects in the four-dimensional Ising model \cite{Lundow:2010en}. From the viewpoint of numerical calculation, it could be possible that there remain some unrevealing aspects in the phase transition of this model and it should be worth trying different approaches other than the Monte Carlo method.

For this purpose we employ the tensor network scheme to investigate the four-dimensional classical Ising model. This scheme has various types of numerical algorithms \cite{Orus:2018dya}, which can be divided into two streams: Hamiltonian approach and Lagrangian one. The latter enables us to evaluate the partition functions directly via tensor network representation. A typical algorithm is the tensor renormalization group (TRG) \cite{Levin:2006jai}, which was originally proposed by Levin and Nave for the two-dimensional Ising model. 
The TRG method has been successfully applied to the two-dimensional field theories with the path-integral formulation in the particle physics \cite{Shimizu:2012zza,Shimizu:2012wfa,Liu:2013nsa,Denbleyker:2013bea,Shimizu:2014uva,Shimizu:2014fsa,Unmuth-Yockey:2014afa,Takeda:2014vwa,Kawauchi:2016dcg,Meurice:2016mkb,Shimizu:2017onf,Kadoh:2018hqq,Sakai:2018xkx,Kadoh:2018tis}. The higher-order TRG (HOTRG) \cite{PhysRevB.86.045139} is an improvement of TRG with the extension to higher dimensions.  One of attractive features in TRG and HOTRG is that we are allowed to directly study the thermodynamic properties; we can systematically increase the system size by repeating the coarse-graining steps in the algorithms. Although earlier studies with HOTRG are restricted to two- and three-dimensional systems \cite{Qin_2013,Yu:2013sbi,Wang_2014,Kawauchi:2015heu,Genzor:2015pua,PhysRevB.93.125115,Kawauchi:2016xng,Sakai:2016tzv,Sakai:2017jwp,Chen:2017ums,Yoshimura:2017jpk,PhysRevE.98.062114,Chen:2018nzs,Kuramashi:2018mmi,NISHINO2019}, including the three-dimensional classical Ising model \cite{PhysRevB.86.045139}, the algorithm itself is readily extended to a four-dimensional lattice. In this paper, we employ the HOTRG method to investigate the phase transition of the classical Ising model on the four-dimensional hypercube. The accuracy of HOTRG is controlled by the bond dimension $\Dcut$, which is varied up to 14 in this study. In order to investigate the phase transition of the model, we measure the internal energy and the magnetization through the evaluation of the tensor network with some impurity located at the center of hypercube.

This paper is organized as follows. In Sec. \ref{sec:TNscheme} we briefly review the HOTRG method and explain an approach to evaluate the internal energy and the magnetization of the Ising model. We present numerical results in Sec. \ref{sec:num} and discuss the properties of the phase transition. Sec. \ref{sec:concl} is devoted to summary and outlook.


\section{HOTRG w/ and w/o impurity}
\label{sec:TNscheme}

The partition function of the four-dimensional ferromagnetic Ising model is given by
\begin{align}
	Z_N = \sum_{\{\sigma=\pm1\}}\left(\prod_{\braket{ij}}T_{\sigma_i\sigma_j}\right)\left(\prod_iV_{\sigma_i}\right)
	\label{eq:pf}
\end{align}
with $T_{\sigma_i\sigma_j}=\e^{\beta\sigma_i\sigma_j}$, $V_{\sigma_i}=\e^{\beta h\sigma_i}$
, where $\sigma_i$ is the two-state classical spin variable on the lattice site $i$, $\braket{ij}$ specifies  the sum over all the nearest-neighboring spin pairs, $\beta$ is the inverse temperature $1/T$ and $h$ is the external magnetic field. The subscript $N$ is the size of a system. Based on the eigenvalue decomposition $T=U\Lambda U^{\T}$, one defines the eight-rank local tensor locating on each lattice site as
\begin{align}
	\mathcal{T}^{(0)}_{i;xx'yy'zz'tt'}=\sum_{\sigma_i}W_{\sigma_ix}W_{\sigma_ix'}W_{\sigma_iy}W_{\sigma_iy'}W_{\sigma_iz}W_{\sigma_iz'}W_{\sigma_it}W_{\sigma_it'}V_{\sigma_i}
\end{align}
where $W=U\sqrt{\Lambda}$. The indices of these tensors are called bond indices. Now, we obtain the tensor network representation of Eq. \eqref{eq:pf} as
\begin{align}
	Z_N=\Tr\prod_{i=1}^{N}\mathcal{T}_i^{(0)},
	\label{eq:TNrep}
\end{align}
where we assume the periodic boundary condition and the right-hand side means all the bond indices are contracted so as to restore the model defined on the four-dimensional hypercube. One way to evaluate Eq.~\eqref{eq:TNrep} is HOTRG with the use of the higher-order singular value decomposition (HOSVD) \cite{PhysRevB.86.045139}. In the HOTRG procedure, nearest two local tensors along the $x$-, $y$-, $z$- and $t$-directions are mapped to the coarse-grained one sequencially. Hence the lattice size is reduced by a factor of 2 after each step of coarse-graining. After repeating $n$ steps of coarse-graining, one obtains the partition function with the system size of $N=2^n$; that is,
\begin{align}
	Z_N\approx\Tr\mathcal{T}_{i=1}^{(n)}.
\end{align}
The right-hand side is again the sum over all the bond indices so as to restore the structure of the four-dimensional lattice model with the periodic boundary condition and this is easily done by defining the trace of the coarse-grained tensor as
\begin{align}
	\Tr\mathcal{T}_{i=1}^{(n)}=\sum_{x,y,z,t}\mathcal{T}^{(n)}_{1;xxyyzztt}.
	\label{eq:tTr}
\end{align}

There are two ways to evaluate the expectation values such as the internal energy and the magnetization. One is the numerical differentiation with respect to $\beta$ and $h$. The other is the direct evaluation of the expectation value using the corresponding tensor network representation. For instance, we can obtain the internal energy through the evaluation of the nearest-neighbor local energy term $\Braket{\sigma_i\sigma_j}$ with the HOTRG method as follows. We first define the additional local tensor as
\begin{align}
	\mathcal{S}^{(0)}_{i;xx'yy'zz'tt'}=\sum_{\sigma_i}\sigma_iW_{\sigma_ix}W_{\sigma_ix'}W_{\sigma_iy}W_{\sigma_iy'}W_{\sigma_iz}W_{\sigma_iz'}W_{\sigma_it}W_{\sigma_it'}V_{\sigma_i}.
\end{align}
With the use of this local tensor, the tensor network representation for the local energy is given by
\begin{align}
	\Braket{\sigma_i\sigma_j}=\Tr\left[\mathcal{S}_i^{(0)}\mathcal{S}_j^{(0)}\prod_{k\neq i,j}\mathcal{T}_k^{(0)}\right]/Z_N,
	\label{eq:impTN}
\end{align}
where $\mathcal{S}_{i,j}^{(0)}$ represent the tensors on the lattice sites $i$ and $j$, respectively, and $\mathcal{T}_k^{(0)}$ is for the rest of $N-2$ sites. Since the numerator looks as if it contains two impurities, we call $\mathcal{S}_i^{(0)}$ impure tensor and $\mathcal{T}_k^{(0)}$ pure tensor. The denominator is evaluated by the plain HOTRG method. In order to coarse-grain the impure tensor network of Eq.~\eqref{eq:impTN}, we assume the local energy term is fixed at the center of lattice ($i=1$, $j=2=1+{\hat y}$ with ${\hat y}$ the unit vector in $y$-direction) during the HOTRG calculation. At the first step, we define the coarse-grained impure tensor $\mathcal{S}_1^{(1)}$ by contracting two initial impurities. For simplicity we give the corresponding expression in the two-dimensional case; 
\begin{align}
	\mathcal{S}^{(1)}_{1;xx'yy'}=\sum_{\alpha,x_1,x'_1,x_2,x'_2}U^{(1)}_{xx_1\otimes x_2}\mathcal{S}^{(0)}_{1;x_1x'_1y\alpha}\mathcal{S}^{(0)}_{2;x_2x'_2\alpha y'}U^{(1)}_{x'x'_1\otimes x'_2},
	\label{eq:SS}
\end{align}
where $U^{(1)}$ is a block-spin transformation determined within the original algorithm of HOTRG \cite{PhysRevB.86.045139}. In the following steps, $\mathcal{S}^{(n+1)}_1$ is defined by the combination of $\mathcal{S}_1^{(n)}$ and $\mathcal{T}_2^{(n)}$. We again show the corresponding formula in two-dimensional case for simplicity;
\begin{align}
	\mathcal{S}^{(n+1)}_{1;xx'yy'}=\sum_{\alpha,x_1,x'_1,x_2,x'_2}U^{(n+1)}_{xx_1\otimes x_2}\mathcal{S}^{(n)}_{1;x_1x'_1y\alpha}\mathcal{T}^{(n)}_{2;x_2x'_2\alpha y'}U^{(n+1)}_{x'x'_1\otimes x'_2}.
	\label{eq:ST}
\end{align}
Finally, the local energy is approximately given by
\begin{align}
	\Braket{\sigma_i\sigma_j}\approx\frac{\Tr\mathcal{S}_1^{(n)}}{\Tr\mathcal{T}_1^{(n)}}.
	\label{eq:energy}
\end{align}
The meaning of the trace is the same as in Eq.~\eqref{eq:tTr}. Since the original model has the translational invariance, $\Braket{\sigma_i\sigma_j}\times d$, where $d$ is the dimensionality, should give the absolute value of internal energy. 

The one-point function $\Braket{\sigma_i}$ to measure the magnetization is also evaluated in the same way. In this case, we are allowed to apply the four-dimensional counterpart of Eq.~\eqref{eq:ST} from the first coarse-graining step because the initial expression of $\Braket{\sigma_i}$ has the form
\begin{align}
	\Braket{\sigma_i}=\Tr\left[\mathcal{S}_i^{(0)}\prod_{k\neq i}\mathcal{T}_k^{(0)}\right]/Z_N,
	\label{eq:impTNspin}
\end{align}
where $\mathcal{S}_i^{(0)}$ locates only on the lattice site $i$ ($i=1$) and $k$ runs the rest of $N-1$ sites. After sufficient iterations, $\Braket{\sigma_i}$ is evaluated by
\begin{align}
	\Braket{\sigma_i}\approx\frac{\Tr\mathcal{S}_1^{(n)}}{\Tr\mathcal{T}_1^{(n)}}.
	\label{eq:spin}
\end{align}
Thanks to the translational invariance, $\Braket{\sigma_i}$ directly corresponds to the spatial average of the Ising spin. Note that Eqs.~\eqref{eq:energy} and ~\eqref{eq:spin} have the same expression, but they are evaluated by different coarse-graining procedures. 

In Ref.~\cite{PhysRevB.86.045139}, computational costs and memory space requirements in two- and three-dimensional HOTRG are given. Computational costs are $\mathcal{O}(\Dcut^{7})$ and $\mathcal{O}(\Dcut^{11})$ and memory space requirements are $\mathcal{O}(\Dcut^{4})$ and $\mathcal{O}(\Dcut^{6})$, respectively. In straightforward expansion of the HOTRG algorithm in Ref.~\cite{PhysRevB.86.045139} to four dimensions, the computational cost is $\mathcal{O}(\Dcut^{15})$ and the memory space requirement is $\mathcal{O}(\Dcut^{8})$. In our implementation, the computational cost in each process is $\mathcal{O}(\Dcut^{13})$ and the memory space requirement in each process is $\mathcal{O}(\Dcut^{7})$. Reduction of the order of computational cost is achieved by using $\Dcut^2$ processes. This implementation is basically based on an idea which will be shown in Ref. \cite{Yamashita:} .We have carried out a detailed measurement of the internal energy and the magnetization with $\Dcut=13$ employing the fine resolution of the temperature $\Delta T=6.25\times 10^{-6}$ around the transition temperature. We have repeated the calculation with $\Dcut=14$ to confirm the qualitative features obtained with $\Dcut=13$. But, in this case the temperature resolution remains coarser as $\Delta T=3.0\times 10^{-5}$ due to the computational cost. In the following we focus on the results with $\Dcut=13$. 

As found in Sec.~\ref{sec:num} all the measured physical quantities seem to lose the volume dependence beyond $n\approx 30$ so that the lattice size of $N=2^{40}=1024^4$ is large enough to be taken as the thermodynamic limit.


\section{Numerical results}
\label{sec:num}

\begin{figure}[t]
	\centering
	\includegraphics[scale=0.8, angle=-90]{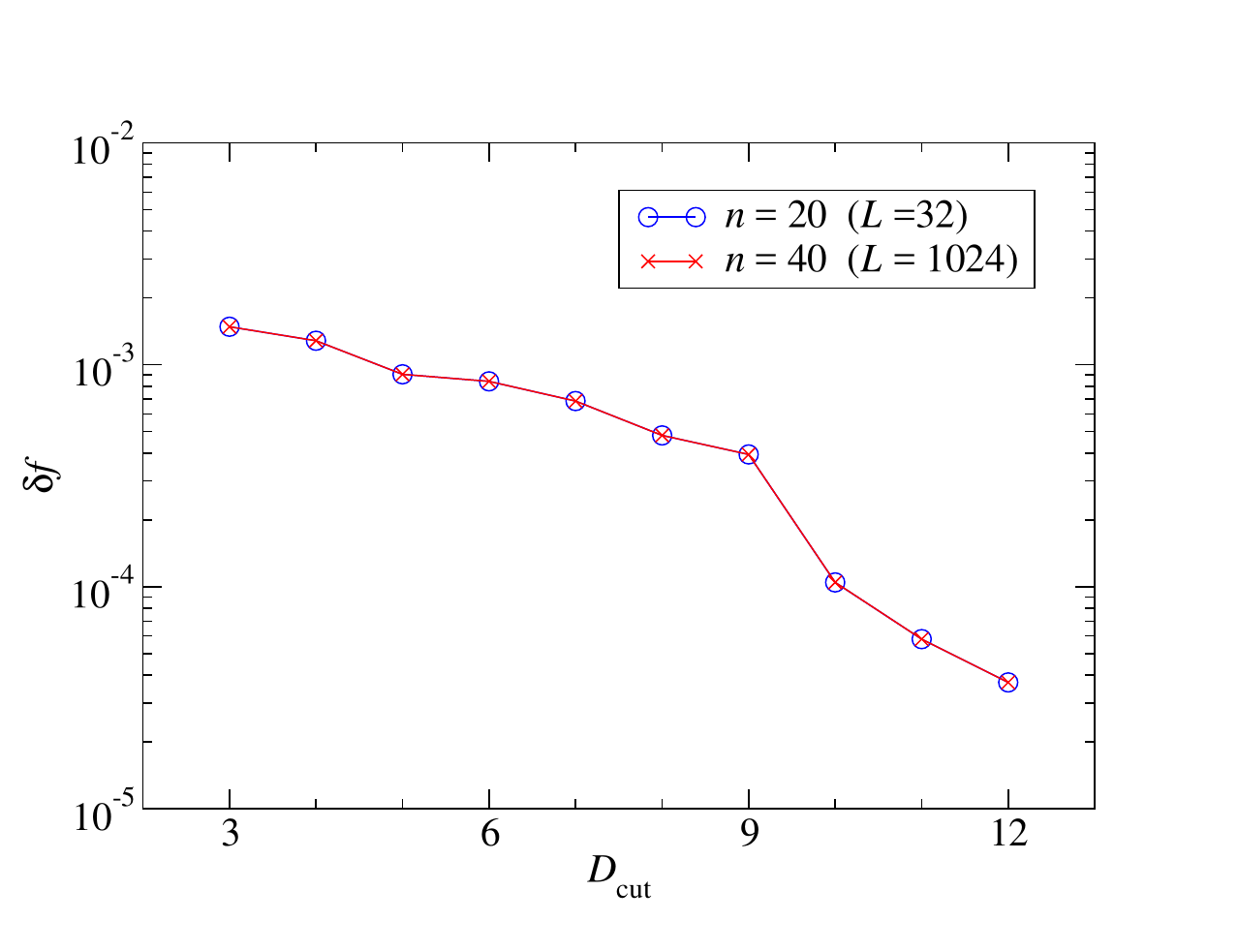}
	\caption{
		Convergence behavior of $\ln Z_N$ as a function of bond dimension $\Dcut$ at $T=6.64250$ in the vicinity of the transition temperature. $L$ is a linear extent of the lattice defined as $N=L^4$.
	}
	\label{fig:dlnz}
\end{figure}

We first evaluate the free energy with the plain HOTRG method. The convergence behavior is investigated by defining the following quantity;
\begin{align}
	\delta f=\left|\frac{\ln Z_N(\Dcut)-\ln Z_N(\Dcut=13)}{\ln Z_N(\Dcut=13)}\right|.
\end{align}
Figure \ref{fig:dlnz} shows a typical convergence behavior of $\ln Z_N$ in the vicinity of the transition temperature. We observe that $\delta f$ decreases monotonically as a function of $\Dcut$.

\begin{figure}[t]
	\centering
	\includegraphics[scale=0.8, angle=-90]{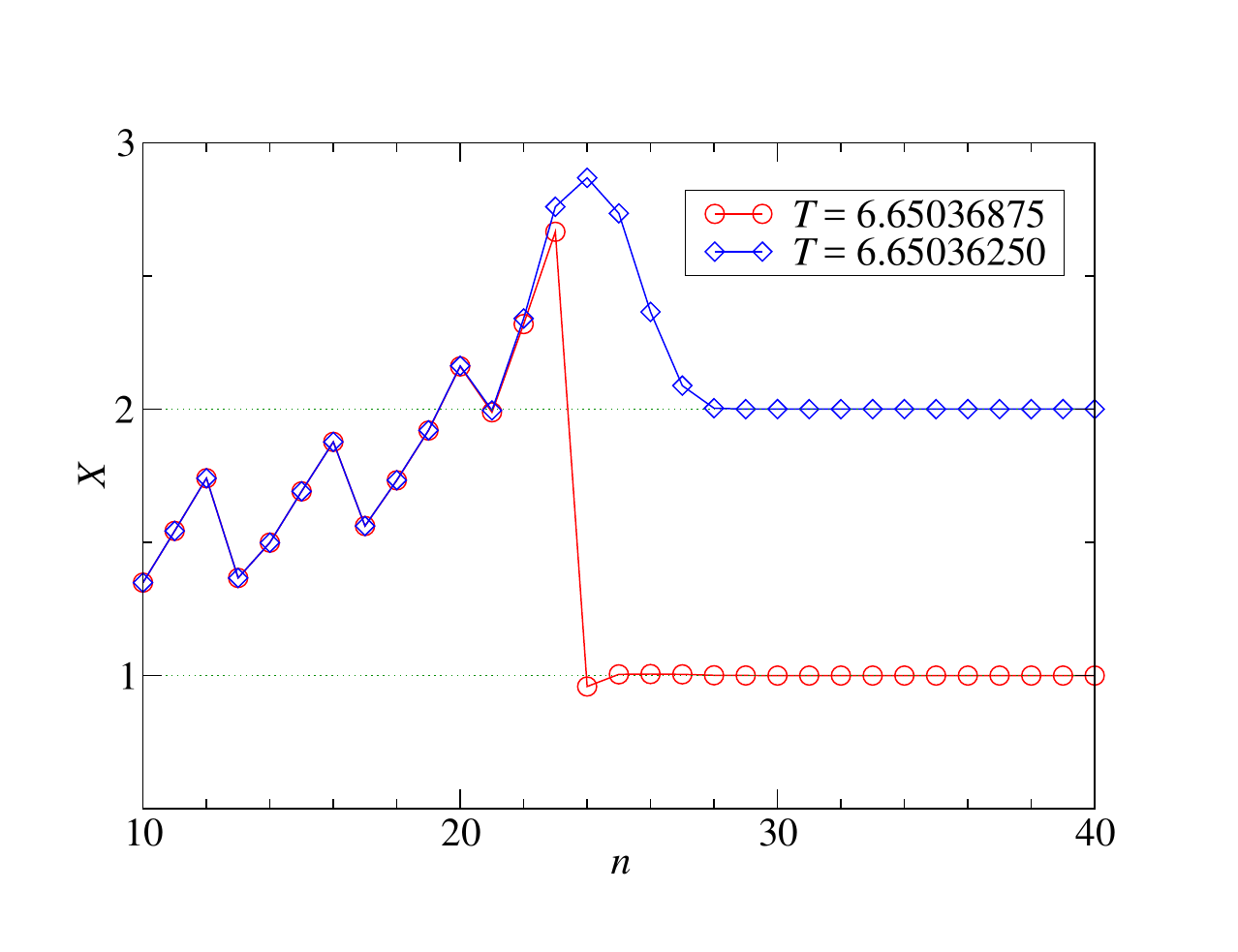}
	\caption{
		$X^{(n)}$ at the $n$-th coarse-graining step with $\Dcut=13$. Red line corresponds to the disordered phase and blue one does to the ordered one.
		}
	\label{fig:x}
\end{figure}

\begin{figure}[t]
	\centering
	\includegraphics[scale=0.8, angle=-90]{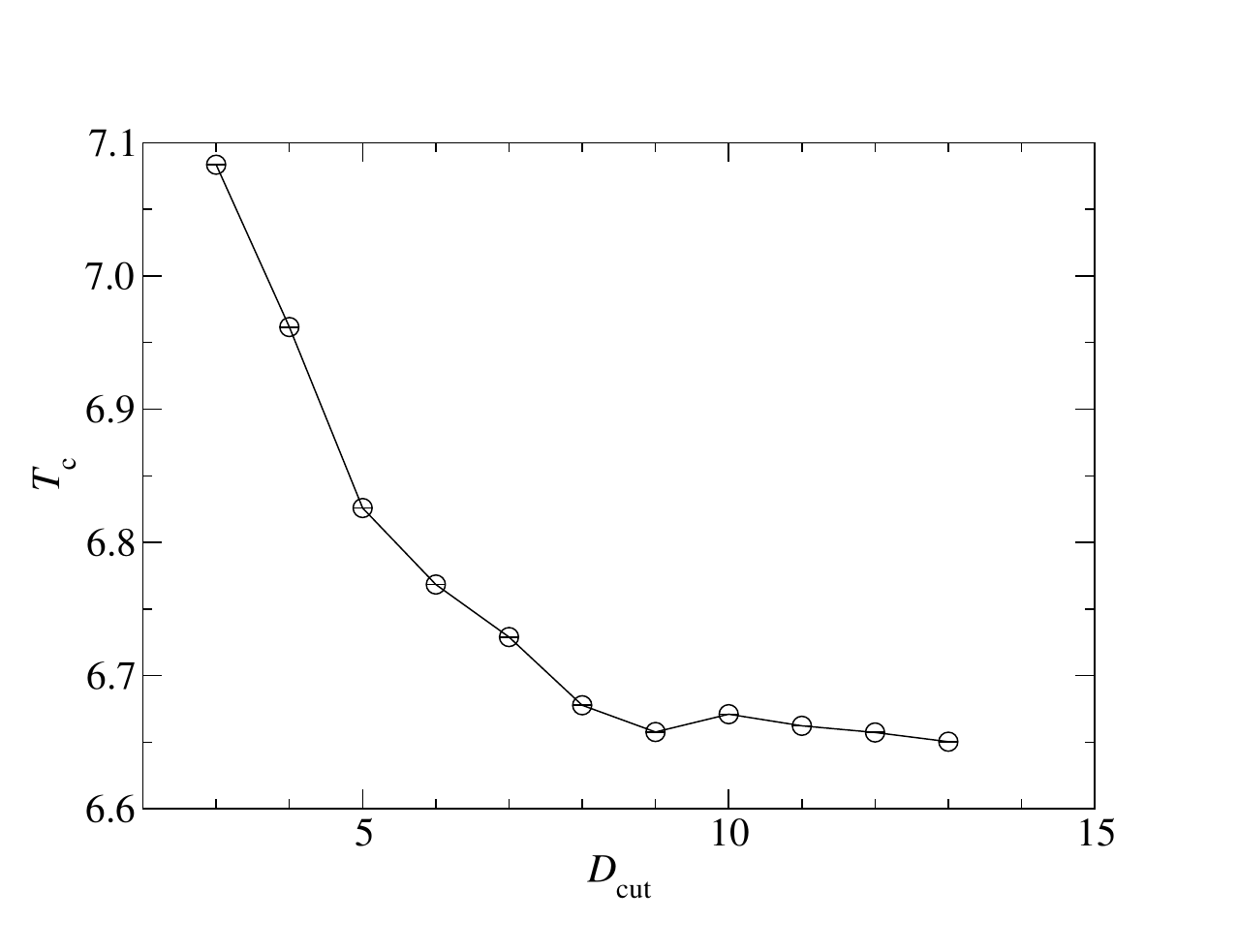}
	\caption{
		Transition temperature as a function of bond dimension. Error bars are within symbols.
		}
	\label{fig:Tc}
\end{figure}

We now turn to the determination of the transition temperature. Let us assume that one has just obtained the coarse-grained tensor $\mathcal{T}^{(n)}_{1;xx'yy'zz'tt'}$ whose coarse-graining direction was $t$-direction. Choosing a $\Dcut\times\Dcut$ matrix as
\begin{align}
	A^{(n)}_{tt'}=\sum_{x,y,z}\mathcal{T}^{(n)}_{1;xxyyzztt'},
	\label{eq:matA}
\end{align}
we calculate the following quantity;
\begin{align}
	X^{(n)}=\frac{\left(\Tr A^{(n)}\right)^2}{\Tr \left(A^{(n)}\right)^2},
\label{eq:x}
\end{align}
which counts the number of the largest singular value of $A^{(n)}$. This is an indicator of the symmetry-breaking \cite{PhysRevB.80.155131}. We calculate $X^{(n)}$ iteratively until it converges. A typical convergence behavior of $X^{(n)}$ is shown in Fig.~\ref{fig:x}. Notice that we sequentially redefine $A^{(n)}$ corresponding to the direction of coarse-graining in the practical calculation. Figure \ref{fig:Tc} shows the transition temperature $\Tc$ as a function of $\Dcut$. The error bars, provided by the temperature resolution, are all smaller than the corresponding symbols. Since $\Tc(\Dcut)$ is estimated by $X^{(n)}$ with sufficiently large $n$, typically beyond $n=30$, there remains little finite-volume effect. In this work, we have obtained $\Tc(\Dcut=13)=6.650365(5)$ on $1024^4$ lattice. The recent Monte Carlo study \cite{PhysRevE.80.031104} obtained $\beta_{\mathrm{c}}=0.1496947(5)$ corresponding to $\Tc=6.68026(2)$, which shows a slight deviation from our result with  HOTRG up to $\Dcut=13$. 
Note that the value of $\Tc$ in Ref.~\cite{PhysRevE.80.031104} was obtained by the infinite-volume extrapolation using the results on relatively small lattices with $L^4\le 80^4$.

\begin{figure}[t]
	\centering
	\includegraphics[scale=0.8, angle=-90]{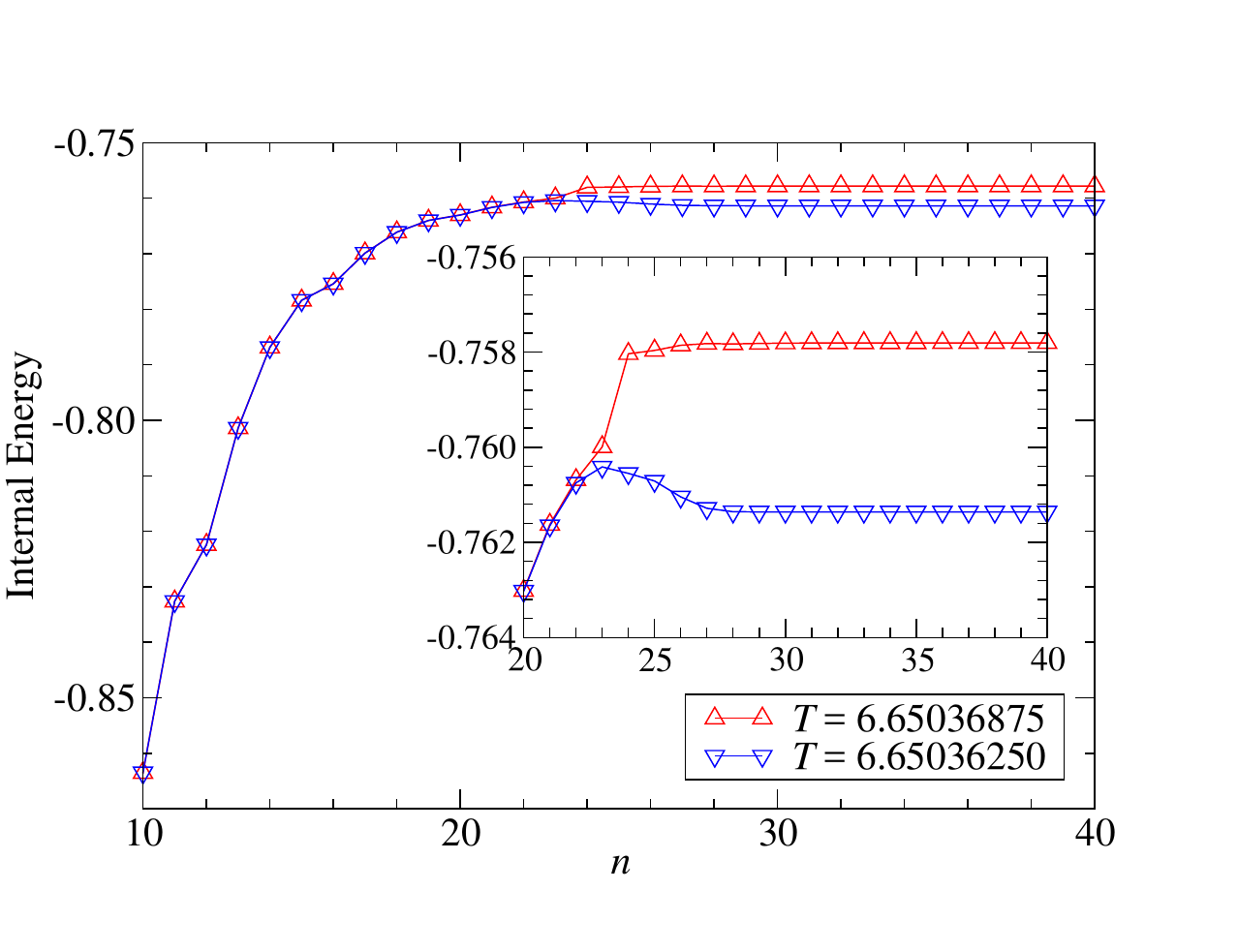}
	\caption{
		Internal energy at the $n$-th coarse-graining step with $\Dcut=13$. Red line corresponds to the disordered phase and blue one does to the ordered phase. Inset graph magnifies the $n$ dependence beyond $n=20$.
		}
	\label{fig:d14ent}
\end{figure}

\begin{figure}[t]
	\centering
	\includegraphics[scale=0.8, angle=-90]{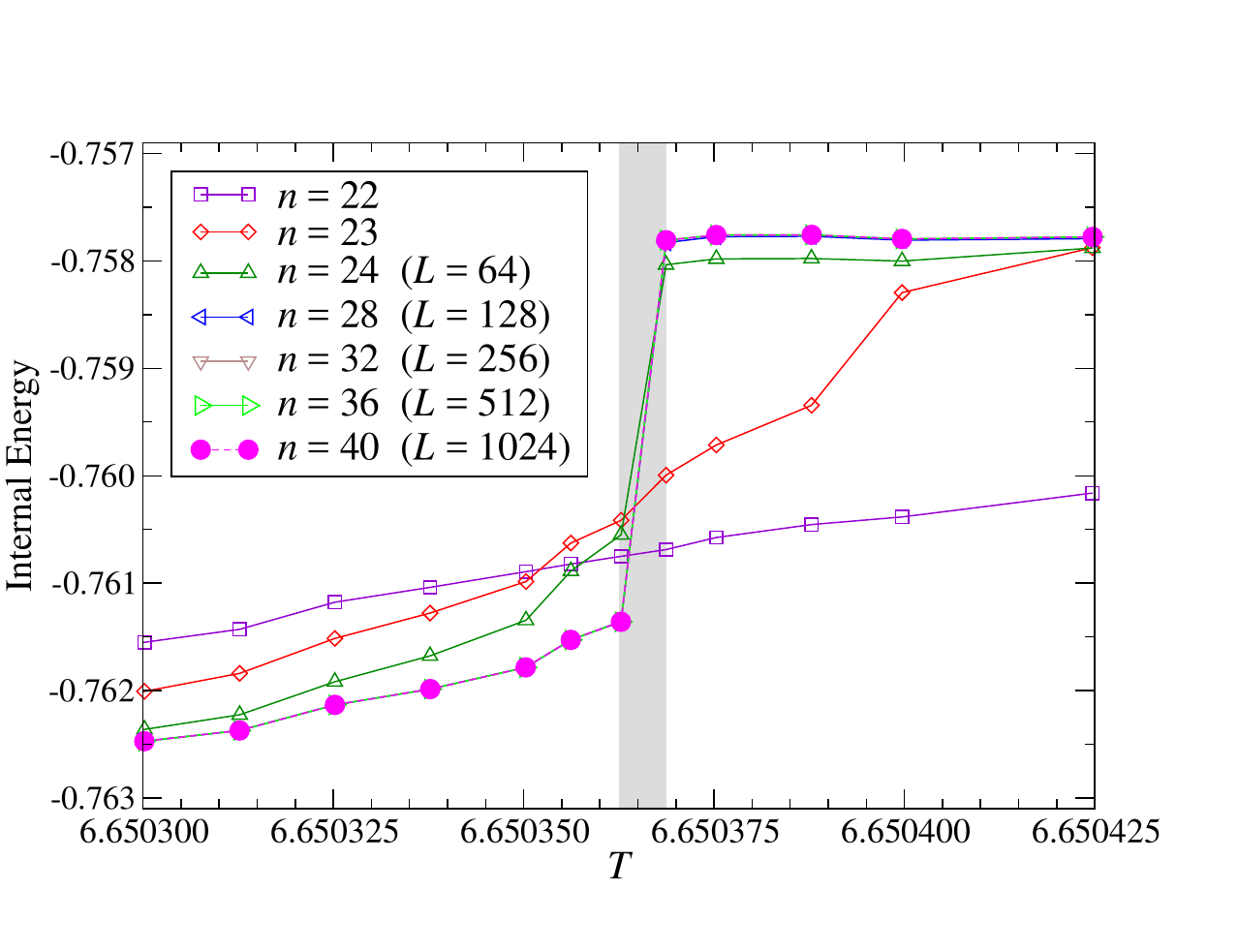}
	\caption{
		Internal energy as a function of temperature for various lattice size with $\Dcut=13$. $\Tc(\Dcut=13)$ estimated by $X^{(n)}$ of Eq.~\eqref{eq:x} is within the gray band.
		}
	\label{fig:etN}
\end{figure}

Let us move on to the evaluation of the internal energy, which can be obtained by numerical differentiation or the coarse-graining of the impure network of Eq.~\eqref{eq:impTN}. We have compared both methods varying the temperature resolution and found that the latter successfully keeps the numerical accuracy as the resolution becomes finer. In the following, we show the results with the impure tensor method. Figure \ref{fig:d14ent} traces the volume dependence of the internal energy with $\Dcut=13$. The converging behavior toward the thermodynamic limit is clearly observed. Since the system size $N$ is given by $2^{n}$, a hypercubic structure is restored in the condition of $n~\mathrm{mod}4=0$. Figure \ref{fig:etN} shows the internal energy as a function of temperature for various lattice sizes with $\Dcut=13$. In the case of $n\ge 24$ ($L\ge 64$), a finite jump emerges with mutual crossings of curves between different volumes around the transition temperature. These are characteristic features of the first-order phase transition as discussed in Ref.~\cite{Fukugita1990}. The similar volume dependence and a finite jump at $L\ge 64$ have been also confirmed in case of $\Dcut=14$. The numerical value of the finite jump $\Delta E(\Dcut=13)$ in the infinite-volume limit is
\begin{align*}
	\Delta E(\Dcut=13)=0.0034(5),
\end{align*}
which is obtained by the linear extrapolation toward the transition temperature both from the low and high temperature regions. The resolution of the temperature at the boundary between the two phases is $\Delta T=6.25\times10^{-6}$. 

\begin{figure}[h]
	\centering
	\includegraphics[scale=0.8, angle=-90]{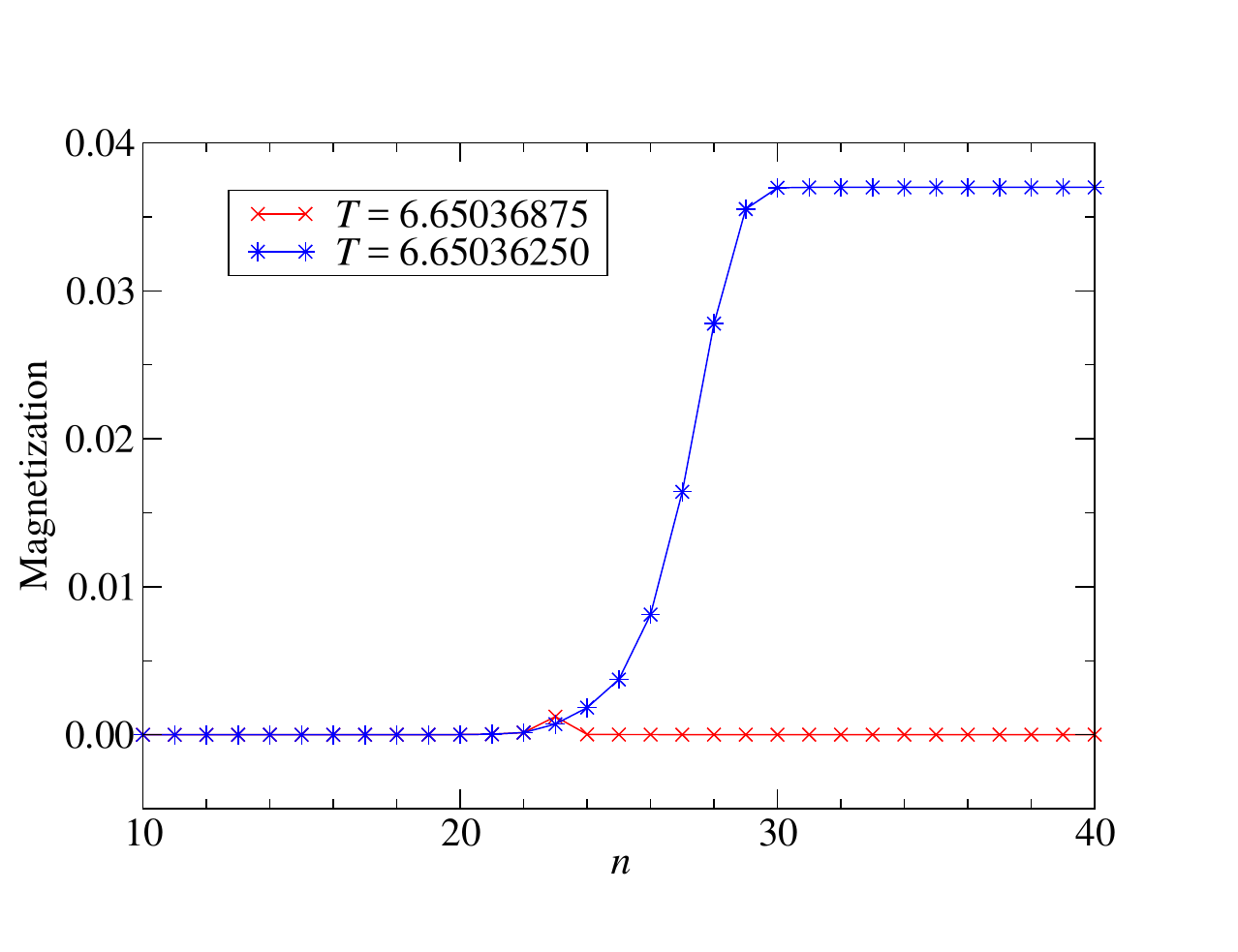}
	\caption{
		Magnetization at the $n$-th coarse-graining step with $\Dcut=13$ and $h=1.0\times10^{-9}$. Red line corresponds to the disordered phase and blue one does to the ordered phase.
		}
	\label{fig:spinnt}
\end{figure}

\begin{figure}[h]
	\centering
	\includegraphics[scale=0.8, angle=-90]{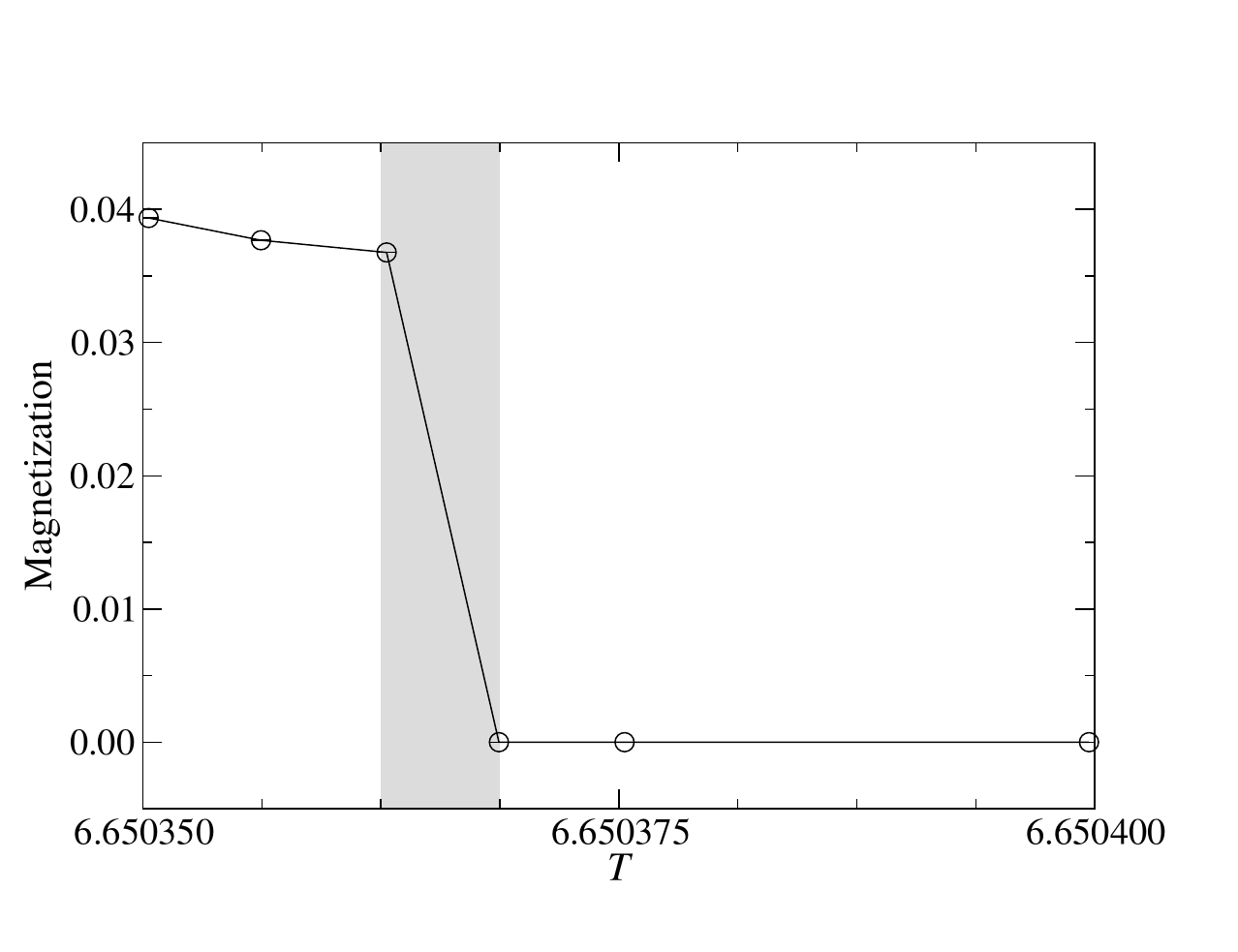}
	\caption{
		Spontaneous magnetization in the thermodynamic limit with $\Dcut=13$. Error bars, provided by extrapolation, are within symbols. $\Tc(\Dcut=13)$ estimated by $X^{(n)}$ of Eq.~\eqref{eq:x} is within the gray band.
		}
	\label{fig:mag}
\end{figure}

We also investigate the spontaneous magnetization, which is an order parameter to detect the symmetry-breaking phase. Figure \ref{fig:spinnt} shows a typical volume dependence of magnetization toward the thermodynamic limit. We have evaluated $\Braket{\sigma_i}$ with $h=1.0\times10^{-9}$ and $2.0\times10^{-9}$ at each temperature and coarse-graining step. After taking the infinite-volume limit we extrapolate the value of $\Braket{\sigma_i}$ toward the $h\to0$ limit. Figure  \ref{fig:mag} shows the resulting spontaneous magnetization as a function of temperature. The transition temperature is consistent with both estimates by $X^{(n)}$ and the internal energy. We have observed a finite jump in the magnetization, whose numerical value is obtained by the linear extrapolation toward the transition temperature both from the low and high temperature regions;
\begin{align*}
\Delta m(\Dcut=13)=0.037(2).
\end{align*}
The resolution of the temperature at the boundary between the two phases is again $\Delta T=6.25\times10^{-6}$. Note that we have tried several choices of the external field  other than $h=\mathcal{O}(10^{-9})$ and confirmed that the behavior of the magnetization is robust against the change of the magnitude of $h$. 

 \section{Summary and outlook}
 \label{sec:concl}
 
We have analized the phase transition of the four-dimensional ferromagnetic Ising model employing HOTRG on $L^4 \le 1024^4$ lattices. The transition temperature is successfully determined by measuring the degeneracy of the largest singular value of the pure tensor. We have also investigated the temperature dependence of the internal energy and magnetization with the impure tensor method. We have found a finite jump for the internal energy together with mutual crossings of curves between different volumes around the transition temperature. A finite jump is also observed in the magnetization. These are characteristic features of the first-order phase transition. The numerical results obtained by the impure tensor method are consistent with the weak first-order phase transition. The resulting estimate for the transition temperature in the thermodynamic limit shows a slight deviation from the recent Monte Carlo prediction \cite{PhysRevE.80.031104} obtained from the infinite-volume extrapolation of the data on relatively small lattices up to $80^4$. 

In future investigation, the HOTRG calculation with $\Dcut>14$ should allow us to achieve a direct and essential improvement of this study. Our impure tensor method can be also improved by considering all the patterns of coarse-graining for the network including some impurities \cite{Morita:2018tpw}. Another possible approach is to develop the best optimization of the Frobenius norm of impure tensor, which would be a realistic way to improve our impure tensor algorithm from the viewpoint of the computational cost of HOTRG. 

\vspace*{+2mm}
    
\begin{acknowledgments}
Numerical calculation for the present work was carried out with the COMA (PACS-IX) computer under the ``Interdisciplinary Computational Science Program" of Center for Computational Sciences, University of Tsukuba and with the Oakforest-PACS system of Joint Center for Advanced High Performance Computing.
This work is supported by the Ministry of Education, Culture, Sports, Science and Technology (MEXT) as ``Exploratory Challenge on Post-K computer (Frontiers of Basic Science: Challenging the Limits)''.
\end{acknowledgments}


\bibliographystyle{JHEP}
\bibliography{trg,4dising,hotrg,forthispaper}

\providecommand{\href}[2]{#2}\begingroup\raggedright\begin{thebibliography}{10}

\bibitem{Aizenman:1981zz}
M.~Aizenman, \emph{{Proof of the Triviality of ${\phi_d}^4$ Field Theory and
  Some Mean-Field Features of Ising Models for $d>4$}},
  \href{https://doi.org/10.1103/PhysRevLett.47.886}{\emph{Phys. Rev. Lett.}
  {\bfseries 47} (1981) 886}.

\bibitem{Aizenman:1982ze}
M.~Aizenman, \emph{{Geometric Analysis of $\phi^4$ Fields and Ising Models.
  Parts I and II}}, \href{https://doi.org/10.1007/BF01205659}{\emph{Commun.
  Math. Phys.} {\bfseries 86} (1982) 1}.

\bibitem{PhysRevB.7.248}
F.~J. Wegner and E.~K. Riedel, \emph{{Logarithmic Corrections to the
  Molecular-Field Behavior of Critical and Tricritical Systems}},
  \href{https://doi.org/10.1103/PhysRevB.7.248}{\emph{Phys. Rev. B} {\bfseries
  7} (1973) 248}.

\bibitem{Wilson:1973jj}
K.~G. Wilson and J.~B. Kogut, \emph{{The Renormalization group and the
  $\varepsilon$ expansion}},
  \href{https://doi.org/10.1016/0370-1573(74)90023-4}{\emph{Phys. Rept.}
  {\bfseries 12} (1974) 75}.

\bibitem{Luscher:1987ay}
M.~L{\"u}scher and P.~Weisz, \emph{{Scaling Laws and Triviality Bounds in the
  Lattice $\phi^4$ Theory (I) One Component Model in the Symmetric Phase}},
  \href{https://doi.org/10.1016/0550-3213(87)90177-5}{\emph{Nucl. Phys.}
  {\bfseries B290} (1987) 25}.

\bibitem{Luscher:1987ek}
M.~L{\"u}scher and P.~Weisz, \emph{{Scaling Laws and Triviality Bounds in the
  Lattice $\phi^4$ Theory (II) One Component Model in the Phase with
  Spontaneous Symmetry Breaking}},
  \href{https://doi.org/10.1016/0550-3213(88)90228-3}{\emph{Nucl. Phys.}
  {\bfseries B295} (1988) 65}.

\bibitem{Luscher:1988uq}
M.~L{\"u}scher and P.~Weisz, \emph{{Scaling Laws and Triviality Bounds in the
  Lattice $\phi^4$ Theory (III) N Component Model}},
  \href{https://doi.org/10.1016/0550-3213(89)90637-8}{\emph{Nucl. Phys.}
  {\bfseries B318} (1989) 705}.

\bibitem{Huang:1988hu}
K.~Huang, \emph{{Triviality of the Higgs Field}},
  \href{https://doi.org/10.1142/S0217751X89000479}{\emph{Int. J. Mod. Phys.}
  {\bfseries A4} (1989) 1037}.

\bibitem{Jansen:1988cw}
K.~Jansen, T.~Trappenberg, I.~Montvay, G.~Munster and U.~Wolff, \emph{{Broken
  Phase of the Four-dimensional Ising Model in a Finite Volume}},
  \href{https://doi.org/10.1016/0550-3213(89)90233-2}{\emph{Nucl. Phys.}
  {\bfseries B322} (1989) 698}.

\bibitem{Frick:1989gw}
C.~Frick, K.~Jansen, J.~Jersak, I.~Montvay, P.~Seuferling and G.~Munster,
  \emph{{Numerical Simulation of the Scalar Sector of the Standard Model in the
  Symmetric Phase}},
  \href{https://doi.org/10.1016/0550-3213(90)90218-3}{\emph{Nucl. Phys.}
  {\bfseries B331} (1990) 515}.

\bibitem{Shrock:2014zca}
R.~Shrock, \emph{{Question of an ultraviolet zero of the beta function of the
  $\lambda (\vec{\phi}^2)^2_4$ theory}},
  \href{https://doi.org/10.1103/PhysRevD.90.065023}{\emph{Phys. Rev.}
  {\bfseries D90} (2014) 065023}
  [\href{https://arxiv.org/abs/1408.3141}{{\ttfamily 1408.3141}}].

\bibitem{Shrock:2016hqn}
R.~Shrock, \emph{{Study of the six-loop beta function of the $\lambda\phi^4_4$
  theory}}, \href{https://doi.org/10.1103/PhysRevD.94.125026}{\emph{Phys. Rev.}
  {\bfseries D94} (2016) 125026}
  [\href{https://arxiv.org/abs/1610.03733}{{\ttfamily 1610.03733}}].

\bibitem{Shrock:2017zuk}
R.~Shrock, \emph{{Study of the question of an ultraviolet zero in the six-loop
  beta function of the O($N$) $\lambda |\vec \phi|^4$ theory}},
  \href{https://doi.org/10.1103/PhysRevD.96.056010}{\emph{Phys. Rev.}
  {\bfseries D96} (2017) 056010}
  [\href{https://arxiv.org/abs/1707.06248}{{\ttfamily 1707.06248}}].

\bibitem{Kenna:1992np}
R.~Kenna and C.~B. Lang, \emph{{Renormalization group analysis of finite size
  scaling in the $\phi^4_4$ model}},
  \href{https://doi.org/10.1016/0550-3213(93)90068-Z,
  10.1016/0550-3213(94)90063-9}{\emph{Nucl. Phys.} {\bfseries B393} (1993) 461}
  [\href{https://arxiv.org/abs/hep-lat/9210009}{{\ttfamily hep-lat/9210009}}].

\bibitem{Kenna:2004cm}
R.~Kenna, \emph{{Finite size scaling for $O(N)$ $\phi^4$-theory at the upper
  critical dimension}},
  \href{https://doi.org/10.1016/j.nuclphysb.2004.05.012}{\emph{Nucl. Phys.}
  {\bfseries B691} (2004) 292}
  [\href{https://arxiv.org/abs/hep-lat/0405023}{{\ttfamily hep-lat/0405023}}].

\bibitem{PhysRevB.22.4481}
H.~W.~J. Bl{\"o}te and R.~H. Swendsen, \emph{{Critical behavior of the
  four-dimensional Ising model}},
  \href{https://doi.org/10.1103/PhysRevB.22.4481}{\emph{Phys. Rev. B}
  {\bfseries 22} (1980) 4481}.

\bibitem{SanchezVelasco:1987ah}
E.~Sanchez-Velasco, \emph{{A finite-size scaling study of the 4D Ising model}},
  \href{https://doi.org/10.1088/0305-4470/20/14/041}{\emph{J. Phys.} {\bfseries
  A20} (1987) 5033}.

\bibitem{Bittner:2002pk}
E.~Bittner, W.~Janke and H.~Markum, \emph{{Ising spins coupled to a
  four-dimensional discrete Regge skeleton}},
  \href{https://doi.org/10.1103/PhysRevD.66.024008}{\emph{Phys. Rev.}
  {\bfseries D66} (2002) 024008}
  [\href{https://arxiv.org/abs/hep-lat/0205023}{{\ttfamily hep-lat/0205023}}].

\bibitem{PhysRevE.80.031104}
P.~H. Lundow and K.~Markstr{\"o}m, \emph{{Critical behavior of the Ising model
  on the four-dimensional cubic lattice}},
  \href{https://doi.org/10.1103/PhysRevE.80.031104}{\emph{Phys. Rev. E}
  {\bfseries 80} (2009) 031104}.

\bibitem{Lundow:2010en}
P.~H. Lundow and K.~Markstr{\"o}m, \emph{{Non-vanishing boundary effects and
  quasi-first order phase transitions in high dimensional Ising models}},
  \href{https://doi.org/10.1016/j.nuclphysb.2010.12.002}{\emph{Nucl. Phys.}
  {\bfseries B845} (2011) 120}
  [\href{https://arxiv.org/abs/1010.5958}{{\ttfamily 1010.5958}}].

\bibitem{Cea:2005ad}
P.~Cea, M.~Consoli and L.~Cosmai, \emph{{Large logarithmic rescaling of the
  scalar condensate: A Subtlety with substantial phenomenological
  implications}},  \href{https://arxiv.org/abs/hep-lat/0501013}{{\ttfamily
  hep-lat/0501013}}.

\bibitem{Stevenson:2005yn}
P.~M. Stevenson, \emph{{Comparison of conventional RG theory with lattice data
  for the 4d Ising model}},
  \href{https://doi.org/10.1016/j.nuclphysb.2005.09.015}{\emph{Nucl. Phys.}
  {\bfseries B729} (2005) 542}
  [\href{https://arxiv.org/abs/hep-lat/0507038}{{\ttfamily hep-lat/0507038}}].

\bibitem{Balog:2006fs}
J.~Balog, F.~Niedermayer and P.~Weisz, \emph{{Repairing Stevenson's step in the
  4d Ising model}},
  \href{https://doi.org/10.1016/j.nuclphysb.2006.02.026}{\emph{Nucl. Phys.}
  {\bfseries B741} (2006) 390}
  [\href{https://arxiv.org/abs/hep-lat/0601016}{{\ttfamily hep-lat/0601016}}].

\bibitem{Orus:2018dya}
{Or{\'u}s, Rom{\'a}n}, \emph{{Tensor networks for complex quantum systems}},
  \href{https://arxiv.org/abs/1812.04011}{{\ttfamily 1812.04011}}.

\bibitem{Levin:2006jai}
M.~Levin and C.~P. Nave, \emph{{Tensor renormalization group approach to
  two-dimensional classical lattice models}},
  \href{https://doi.org/10.1103/PhysRevLett.99.120601}{\emph{Phys. Rev. Lett.}
  {\bfseries 99} (2007) 120601}
  [\href{https://arxiv.org/abs/cond-mat/0611687}{{\ttfamily
  cond-mat/0611687}}].

\bibitem{Shimizu:2012zza}
Y.~Shimizu, \emph{{Tensor renormalization group approach to a lattice boson
  model}}, \href{https://doi.org/10.1142/S0217732312500356}{\emph{Mod. Phys.
  Lett.} {\bfseries A27} (2012) 1250035}.

\bibitem{Shimizu:2012wfa}
Y.~Shimizu, \emph{{Analysis of the (1+1)-dimensional lattice $\phi^{4}$ model
  using the tensor renormalization group}}, {\emph{Chin. J. Phys.} {\bfseries
  50} (2012) 749}.

\bibitem{Liu:2013nsa}
Y.~Liu, Y.~Meurice, M.~P. Qin, J.~Unmuth-Yockey, T.~Xiang, Z.~Y. Xie et~al.,
  \emph{{Exact Blocking Formulas for Spin and Gauge Models}},
  \href{https://doi.org/10.1103/PhysRevD.88.056005}{\emph{Phys. Rev.}
  {\bfseries D88} (2013) 056005}
  [\href{https://arxiv.org/abs/1307.6543}{{\ttfamily 1307.6543}}].

\bibitem{Denbleyker:2013bea}
A.~Denbleyker, Y.~Liu, Y.~Meurice, M.~P. Qin, T.~Xiang, Z.~Y. Xie et~al.,
  \emph{{Controlling Sign Problems in Spin Models Using Tensor
  Renormalization}},
  \href{https://doi.org/10.1103/PhysRevD.89.016008}{\emph{Phys. Rev.}
  {\bfseries D89} (2014) 016008}
  [\href{https://arxiv.org/abs/1309.6623}{{\ttfamily 1309.6623}}].

\bibitem{Shimizu:2014uva}
Y.~Shimizu and Y.~Kuramashi, \emph{{Grassmann tensor renormalization group
  approach to one-flavor lattice Schwinger model}},
  \href{https://doi.org/10.1103/PhysRevD.90.014508}{\emph{Phys. Rev.}
  {\bfseries D90} (2014) 014508}
  [\href{https://arxiv.org/abs/1403.0642}{{\ttfamily 1403.0642}}].

\bibitem{Shimizu:2014fsa}
Y.~Shimizu and Y.~Kuramashi, \emph{{Critical behavior of the lattice Schwinger
  model with a topological term at $\theta=\pi$ using the Grassmann tensor
  renormalization group}},
  \href{https://doi.org/10.1103/PhysRevD.90.074503}{\emph{Phys. Rev.}
  {\bfseries D90} (2014) 074503}
  [\href{https://arxiv.org/abs/1408.0897}{{\ttfamily 1408.0897}}].

\bibitem{Unmuth-Yockey:2014afa}
J.~F. Unmuth-Yockey, Y.~Meurice, J.~Osborn and H.~Zou, \emph{{Tensor
  renormalization group study of the 2d O(3) model}},
  \href{https://doi.org/10.22323/1.214.0325}{\emph{PoS} {\bfseries LATTICE2014}
  (2014) 325} [\href{https://arxiv.org/abs/1411.4213}{{\ttfamily 1411.4213}}].

\bibitem{Takeda:2014vwa}
S.~Takeda and Y.~Yoshimura, \emph{{Grassmann tensor renormalization group for
  the one-flavor lattice Gross-Neveu model with finite chemical potential}},
  \href{https://doi.org/10.1093/ptep/ptv022}{\emph{PTEP} {\bfseries 2015}
  (2015) 043B01} [\href{https://arxiv.org/abs/1412.7855}{{\ttfamily
  1412.7855}}].

\bibitem{Kawauchi:2016dcg}
H.~Kawauchi and S.~Takeda, \emph{{Phase structure analysis of CP($N$-1) model
  using Tensor renormalization group}},
  \href{https://doi.org/10.22323/1.256.0322}{\emph{PoS} {\bfseries LATTICE2016}
  (2016) 322} [\href{https://arxiv.org/abs/1611.00921}{{\ttfamily
  1611.00921}}].

\bibitem{Meurice:2016mkb}
Y.~Meurice, A.~Bazavov, S.-W. Tsai, J.~Unmuth-Yockey, L.-P. Yang and J.~Zhang,
  \emph{{Tensor RG calculations and quantum simulations near criticality}},
  \href{https://doi.org/10.22323/1.256.0325}{\emph{PoS} {\bfseries LATTICE2016}
  (2016) 325} [\href{https://arxiv.org/abs/1611.08711}{{\ttfamily
  1611.08711}}].

\bibitem{Shimizu:2017onf}
Y.~Shimizu and Y.~Kuramashi, \emph{{Berezinskii-Kosterlitz-Thouless transition
  in lattice Schwinger model with one flavor of Wilson fermion}},
  \href{https://doi.org/10.1103/PhysRevD.97.034502}{\emph{Phys. Rev.}
  {\bfseries D97} (2018) 034502}
  [\href{https://arxiv.org/abs/1712.07808}{{\ttfamily 1712.07808}}].

\bibitem{Kadoh:2018hqq}
D.~Kadoh, Y.~Kuramashi, Y.~Nakamura, R.~Sakai, S.~Takeda and Y.~Yoshimura,
  \emph{{Tensor network formulation for two-dimensional lattice $ \mathcal{N} $
  = 1 Wess-Zumino model}},
  \href{https://doi.org/10.1007/JHEP03(2018)141}{\emph{JHEP} {\bfseries 03}
  (2018) 141} [\href{https://arxiv.org/abs/1801.04183}{{\ttfamily
  1801.04183}}].

\bibitem{Sakai:2018xkx}
R.~Sakai, D.~Kadoh, Y.~Kuramashi, Y.~Nakamura, S.~Takeda and Y.~Yoshimura,
  \emph{{Tensor network study of two dimensional lattice $\phi^{4}$ theory}},
  \href{https://doi.org/10.22323/1.334.0232}{\emph{PoS} {\bfseries LATTICE2018}
  (2018) 232} [\href{https://arxiv.org/abs/1812.00166}{{\ttfamily
  1812.00166}}].

\bibitem{Kadoh:2018tis}
D.~Kadoh, Y.~Kuramashi, Y.~Nakamura, R.~Sakai, S.~Takeda and Y.~Yoshimura,
  \emph{{Tensor network analysis of critical coupling in two dimensional
  $\phi^{4}$ theory}},
  \href{https://doi.org/10.1007/JHEP05(2019)184}{\emph{JHEP} {\bfseries 05}
  (2019) 184} [\href{https://arxiv.org/abs/1811.12376}{{\ttfamily
  1811.12376}}].

\bibitem{PhysRevB.86.045139}
Z.~Y. Xie, J.~Chen, M.~P. Qin, J.~W. Zhu, L.~P. Yang and T.~Xiang,
  \emph{{Coarse-graining renormalization by higher-order singular value
  decomposition}},
  \href{https://doi.org/10.1103/PhysRevB.86.045139}{\emph{Phys. Rev. B}
  {\bfseries 86} (2012) 045139}.

\bibitem{Qin_2013}
M.-P. Qin, J.~Chen, Q.-N. Chen, Z.-Y. Xie, X.~Kong, H.-H. Zhao et~al.,
  \emph{{Partial Order in Potts Models on the Generalized Decorated Square
  Lattice}}, \href{https://doi.org/10.1088/0256-307x/30/7/076402}{\emph{Chinese
  Physics Letters} {\bfseries 30} (2013) 076402}.

\bibitem{Yu:2013sbi}
J.~F. Yu, Z.~Y. Xie, Y.~Meurice, Y.~Liu, A.~Denbleyker, H.~Zou et~al.,
  \emph{{Tensor Renormalization Group Study of Classical XY Model on the Square
  Lattice}}, \href{https://doi.org/10.1103/PhysRevE.89.013308}{\emph{Phys.
  Rev.} {\bfseries E89} (2014) 013308}
  [\href{https://arxiv.org/abs/1309.4963}{{\ttfamily 1309.4963}}].

\bibitem{Wang_2014}
S.~Wang, Z.-Y. Xie, J.~Chen, B.~Normand and T.~Xiang, \emph{{Phase Transitions
  of Ferromagnetic Potts Models on the Simple Cubic Lattice}},
  \href{https://doi.org/10.1088/0256-307x/31/7/070503}{\emph{Chinese Physics
  Letters} {\bfseries 31} (2014) 070503}.

\bibitem{Kawauchi:2015heu}
H.~Kawauchi and S.~Takeda, \emph{{Tensor renormalization group analysis of
  CP($N$-1) model in two dimensions}},
  \href{https://doi.org/10.22323/1.251.0284}{\emph{PoS} {\bfseries LATTICE2015}
  (2016) 284} [\href{https://arxiv.org/abs/1511.00348}{{\ttfamily
  1511.00348}}].

\bibitem{Genzor:2015pua}
J.~Genzor, A.~Gendiar and T.~Nishino, \emph{{Phase transition of the Ising
  model on a fractal lattice}},
  \href{https://doi.org/10.1103/PhysRevE.93.012141}{\emph{Phys. Rev.}
  {\bfseries E93} (2016) 012141}
  [\href{https://arxiv.org/abs/1509.05596}{{\ttfamily 1509.05596}}].

\bibitem{PhysRevB.93.125115}
H.-H. Zhao, Z.-Y. Xie, T.~Xiang and M.~Imada, \emph{{Tensor network algorithm
  by coarse-graining tensor renormalization on finite periodic lattices}},
  \href{https://doi.org/10.1103/PhysRevB.93.125115}{\emph{Phys. Rev. B}
  {\bfseries 93} (2016) 125115}.

\bibitem{Kawauchi:2016xng}
H.~Kawauchi and S.~Takeda, \emph{{Tensor renormalization group analysis of
  CP($N$-1) model}},
  \href{https://doi.org/10.1103/PhysRevD.93.114503}{\emph{Phys. Rev.}
  {\bfseries D93} (2016) 114503}
  [\href{https://arxiv.org/abs/1603.09455}{{\ttfamily 1603.09455}}].

\bibitem{Sakai:2016tzv}
R.~Sakai and S.~Takeda, \emph{{Tensor renormalization group approach to higher
  dimensional fermions}}, \href{https://doi.org/10.22323/1.256.0336}{\emph{PoS}
  {\bfseries LATTICE2016} (2016) 336}.

\bibitem{Sakai:2017jwp}
R.~Sakai, S.~Takeda and Y.~Yoshimura, \emph{{Higher-order tensor
  renormalization group for relativistic fermion systems}},
  \href{https://doi.org/10.1093/ptep/ptx080}{\emph{PTEP} {\bfseries 2017}
  (2017) 063B07} [\href{https://arxiv.org/abs/1705.07764}{{\ttfamily
  1705.07764}}].

\bibitem{Chen:2017ums}
J.~Chen, H.-J. Liao, H.-D. Xie, X.-J. Han, R.-Z. Huang, S.~Cheng et~al.,
  \emph{{Phase transition of the q-state clock model: duality and tensor
  renormalization}},
  \href{https://doi.org/10.1088/0256-307X/34/5/050503}{\emph{Chin. Phys. Lett.}
  {\bfseries 34} (2017) 050503}
  [\href{https://arxiv.org/abs/1706.03455}{{\ttfamily 1706.03455}}].

\bibitem{Yoshimura:2017jpk}
Y.~Yoshimura, Y.~Kuramashi, Y.~Nakamura, S.~Takeda and R.~Sakai,
  \emph{{Calculation of fermionic Green functions with Grassmann higher-order
  tensor renormalization group}},
  \href{https://doi.org/10.1103/PhysRevD.97.054511}{\emph{Phys. Rev.}
  {\bfseries D97} (2018) 054511}
  [\href{https://arxiv.org/abs/1711.08121}{{\ttfamily 1711.08121}}].

\bibitem{PhysRevE.98.062114}
R.~Krcmar, J.~Genzor, Y.~Lee, H.~\ifmmode \check{C}\else
  \v{C}\fi{}en\ifmmode~\check{c}\else \v{c}\fi{}arikov\'a, T.~Nishino and
  A.~Gendiar, \emph{{Tensor-network study of a quantum phase transition on the
  Sierpi\ifmmode \acute{n}\else \'{n}\fi{}ski fractal}},
  \href{https://doi.org/10.1103/PhysRevE.98.062114}{\emph{Phys. Rev. E}
  {\bfseries 98} (2018) 062114}.

\bibitem{Chen:2018nzs}
Y.~Chen, Z.-Y. Xie and J.-F. Yu, \emph{{Phase transitions of the five-state
  clock model on the square lattice}},
  \href{https://doi.org/10.1088/1674-1056/27/8/080503}{\emph{Chin. Phys.}
  {\bfseries B27} (2018) 080503}
  [\href{https://arxiv.org/abs/1804.05532}{{\ttfamily 1804.05532}}].

\bibitem{Kuramashi:2018mmi}
Y.~Kuramashi and Y.~Yoshimura, \emph{{Three-dimensional finite temperature
  Z$_2$ gauge theory with tensor network scheme}},
  \href{https://arxiv.org/abs/1808.08025}{{\ttfamily 1808.08025}}.

\bibitem{NISHINO2019}
A.~G. Jozef~Genzor and T.~Nishino, \emph{{Measurements of magnetization on the
  Sierpi\ifmmode \acute{n}\else \'{n}\fi{}ski carpet}},
  \href{https://arxiv.org/abs/1904.10645}{{\ttfamily 1904.10645}}.

\bibitem{Yamashita:}
T.~Yamashita and T.~Sakurai, \emph{{in preparation}}, .

\bibitem{PhysRevB.80.155131}
Z.-C. Gu and X.-G. Wen, \emph{{Tensor-entanglement-filtering renormalization
  approach and symmetry-protected topological order}},
  \href{https://doi.org/10.1103/PhysRevB.80.155131}{\emph{Phys. Rev. B}
  {\bfseries 80} (2009) 155131}.

\bibitem{Fukugita1990}
M.~Fukugita, H.~Mino, M.~Okawa and A.~Ukawa, \emph{{Finite-size scaling of the
  three-state Potts model on a simple cubic lattice}},
  \href{https://doi.org/10.1007/BF01334757}{\emph{Journal of Statistical
  Physics} {\bfseries 59} (1990) 1397}.

\bibitem{Morita:2018tpw}
S.~Morita and N.~Kawashima, \emph{{Calculation of higher-order moments by
  higher-order tensor renormalization group}},
  \href{https://doi.org/10.1016/j.cpc.2018.10.014}{\emph{Comput. Phys. Commun.}
  {\bfseries 236} (2019) 65}
  [\href{https://arxiv.org/abs/1806.10275}{{\ttfamily 1806.10275}}].

\end{thebibliography}\endgroup

\end{document}